\documentclass[12pt]{iopart}
\eqnobysec
\usepackage{iopams}

\usepackage{graphicx}

\newtheorem{theorem}{Theorem}

\newtheorem{definition}{Definition}

\begin{document}

\title{Quantum control and representation theory}

\author{A. Ibort, J. M. P\'erez-Pardo}

\address{Depto. de Matem\'aticas, Univ. Carlos III de
Madrid, Avda. de la Universidad 30, 28911 Legan\'es, Madrid, Spain.}

\ead{albertoi@math.uc3m.es}

\begin{abstract}
A new notion of controllability for quantum systems that takes advantage of the linear superposition of
quantum states is introduced.   We call such notion von Neumann controllabilty and it is shown that it is strictly weaker than the usual notion
of pure state and operator controlability.  We provide a simple and effective characterization of it by using tools from the theory of unitary representations of Lie groups.   In this sense we are able to approach the problem of control of quantum states from a new perspective, that of the theory of unitary representations of Lie groups.    A few examples of physical interest and the particular instances of compact and nilpotent dynamical
Lie groups are discussed.
\end{abstract}

\vspace{2pc}

\maketitle

\section{Introduction: control of infinite-dimensional quantum systems}
The theory of control of quantum systems has  had a strong influence from
ideas from the classical theory of control.  In fact, from a purely mathe\-matical perspective, a quantum system is nothing but a linear
evolution equation on a vector space, thus the methods and ideas in
classical control apply straightforwardly.    This approach has been very succesful
in tackling control problems for finite-dimensional quantum systems (or finite-dimensional approximations to them) where well-known theorems characterizing the various notions of controllability of classical systems 
have been applied.    For instance, the state controllability of a finite-dimensional quantum system has been
considered from a Lie-theoretical perspective determining necessary and
sufficient conditions for a given dynamical group to act transitively on the set of normalized pure
states (see for instance the recent book \cite{Al08} and references therein).    However such
ideas cannot be straightforwardly extended to the infinite-dimensional situation
for various reasons, among them  the intrinsic and unavoidable analytical difficulties coming from the appearance of unbounded
operators and the complications inherit to infinite-dimensional geometry.

Moreover, the results obtained for finite-dimensional approximations of quantum systems do not extend naively
to the infinite-dimensional case.   For instance, it can be easily checked that the harmonic
oscillator control problem:
\begin{equation}\label{harmonic}
i \frac{\partial \psi}{\partial t} = - \frac{1}{2} \frac{\partial^2 \psi}{\partial q^2} + \left( \frac{1}{2} q^2 -u(t) q\right) \psi ,
\end{equation}
is not controllable (see for instance \cite{Mi04}), however its truncations up to the first $n+1$ eigenstates, energies
between $1/2$ and $n + 1/2$,
whose form is:
$$ i\frac{d\tilde\psi}{dt} = (\tilde{H}_0 + u\tilde{H}_1) \tilde\psi, \quad \tilde\psi \in \mathbb{C}^{n+1} ,$$
is controllable for every $n$ \cite{Ra95}.    A few results on the controllability of the 1D Schr\"odinger equation
have appeared recently \cite{Be05}, \cite{Be06} displaying some of the subtleties of the infinite-dimensional case.  
 It is important to point out that a rigorous approximate controllability theorem has been proved for 
general bilinear systems under a non-rational resonance condition of the spectrum 
of the free hamiltonian by Chambrion {\em et al} \cite{Ch08}.

Other recent contributions to the subject have been focusing on the possibility of circumvent
the technical difficulties that arise in the infinite-dimensional setting. For instance
Bloch {\em et al} \cite{Bl06} provide a set of sufficient conditions to prove the controllability of finite-dimensional systems
coupled to harmonic oscillators
that extend the well-known conditions in finite-dimensional geometry. There are also remarkable the results by Clark 
 \cite{Cl03}
on controllability based on the existence of a common domain of analytic vectors for the control hamiltonians $H_k$.

In this work we take a different approach to the problem of state controllability of quantum systems by
relaxing the notion of (approximate) control and allowing for the linear superposition of states.  
In fact, the linear superposition principle constitutes a fundamental ingredient of quantum physics, therefore it can be exploited 
when addressing the problem of control of quantum systems, contrary to what happens in classical systems.
Moreover the use of the superposition principle has attracted a lot of 
attention since the early years of quantum physics and is becoming a more and more relevant
tool in the manipulation of quantum states.   Sometimes superposed states are called cat states
recalling the famous thought experiment by Schr\"odinger.   Linear superposition of states has been achieved, for both photons and electrons,  for various purposes (see for instance \cite{Ma01}, \cite{Er05} for recent applications to quantum information theory).   We refer in this paper to a few concrete experiments with photons,
in particular Mach-Zehnder interferometer with a Kerr medium, that will help
us to illustrate this new notion of controllability \cite{Ge99}, \cite{Kw92}.

Thus, given a control quantum hamiltonian $H(u)$, we ask for the existence of controls such that the 
target state gets arbitrarily close to a linear superposition of
states, each one obtained evolving from a given initial one by the use of (possibly) different families of control functions. 
We call such notion of controllability von Neumann controllability for reasons that will become obvious
from the discussion to follow, and we will relate such notion of state control to the theory of unitary representations
of the dynamical Lie group of the system.    In fact a necessary condition for the pure state controllability of the system 
is the irreducilibity of the corresponding unitary representation of the dynamical group.
It is also shown that the irreducibility of the unitary representation of the dynamical Lie group is equivalent to von Neumann controllability of the system. Thus if the dynamical group $G$ is compact we are led necessarily to consider finite-dimensional representations.  Consequently genuine state controllable infinite-dimensional systems can only arise if the group $G$ is not compact. Some families of infinite-dimensional representations are well-known, for instance
for nilpotent and solvable groups.    We use the knowledge we have
on such representations to obtain some simple results on the various notions of controllability for such systems.

The plan of this article is as follows. Section \ref{controllability} is devoted to establish the basic
notions of controllability and approximate controllability of quantum systems and to dwell into the differences between the
finite and infinite-dimensional cases.
We introduce the notion
of von Neumann controllability in section \ref{von Neumann} and discuss its relation to the representation theory of
the dynamical Lie group of the system in section \ref{representation}.   Some relations between the various notions of controllability are discussed there
and an explicit characterization of von Neumann controllability is given.
Finally a discussion takes place in section \ref{examples} about various, particularly interesting, cases related to compact and nilpotent groups, as well as the
oscillator algebra and its relation to the Mach-Zehnder-Kerr system discussed in section 3, followed by a summary 
of the paper and an outlook of further developments.

\section{Controllability and approximate controllability for quantum systems}\label{controllability}

We shall consider a quantum system defined on a Hilbert space $\mathcal{H}$ with
hamiltonian operator $H(u)$ that depends on a family of control functions $u(t)$
lying in some class $\mathcal{U}$ of admissible controls.  Tipically the control
hamiltonian will be of affine-linear type, i.e.
$$ H(u) = H_0 + \sum_{k = 1}^r u_k(t) H_k ,$$
where the drift hamiltonian $H_0$ represents the dynamics of the uncontrolled (or ``free'') system, and the
set of controls $\mathcal{U}$ is the space of bounded piecewise constant functions $u(t)$.
Given a family of control functions $u(t)$, if $H(u(t))$ is a self-adjoint operator for $t$ in the interval $[0,T]$, there exists a unique solution to the time-dependent Schr\"odinger equation:

\begin{equation}\label{schro}
 i\hbar \dot{\psi} = H(u(t)) \psi ,
 \end{equation}
with initial state $\psi_0 \in \mathcal{H}$.  We shall denote such solution by $\psi (t; u)$.
Moreover there exist a family of unitary operators $U(t)$ which provide
the quantum evolution of the system on Heisenberg's representation, i.e.
$$ i\hbar \dot{U} = H(u(t)) U , \quad U(0) = I, \quad  U \in U(\mathcal{H}),$$
where $U(\mathcal{H})$ denotes the group of unitary operators on $\mathcal{H}$.

For a given time $t>0$ we define the set of reachable states (or operators) $\mathcal{R}_t (\psi_0)=
\{ \psi = \psi(t;u) \in \mathcal{H} \mid u \in \mathcal{U} \}$  ($\mathcal{R}_t^{op} =
\{ U = U(t;u) \in U(\mathcal{H}) \mid u \in \mathcal{U} \}$), and for a given $T>0$ ($T$ could be $+\infty$) we consider the set
of reachable states (or operators) for all times $0\leq t \leq T$, that is $\mathcal{R}(\psi_0,T) = \bigcup_{0\leq t\leq T} \mathcal{R}_t(\psi_0)$
($\mathcal{R}^{op}(T) = \bigcup_{0\leq t \leq T} \mathcal{R}_t^{op}$).
Then, given a subset $\Sigma\subset \mathcal{H}$ and a time $T>0$, we say that the system is $(\Sigma, T)$-controllable with respect to the state $\psi_0$ if $\Sigma \subset \mathcal{R}(\psi_0,T)$ and we say that the system is $(\Sigma, T)$-controllable if it is $(\Sigma, T)$-controllable with respect to any $\psi_0$. Tipically we will be interested on pure state controllability, this is $\Sigma = S \subset \mathcal{H}$,
where $S$ is the unit sphere made up of all unitary vectors on the Hilbert space $\mathcal{H}$ and $T = +\infty$.   Similarly, we say that the system is operator controllable if $\mathcal{R}^{op}(+\infty) = U(\mathcal{H})$.  

There are various criteria for assessing the problem of controllability in finite-dimensional systems, 
however not much is known in the infinite-dimensional situation.     The most successful methods in
finite dimension for affine quantum control systems are based on the identification of the reachable set with an orbit
of the Lie group associated to the dynamical Lie algebra of the system, i.e. the unique connected and simply connected Lie group $G$
such that its Lie algebra $\mathfrak{g}$ is the Lie algebra generated by the operators $iH_0, \ldots, iH_r$.
Hence a $n$-dimensional affine quantum control system is pure state controllable if the Lie group $G$ acts
transitively on the finite-dimensional sphere $S^{2n-1}$.
It is then a simple task based on classical results on homogeneous spaces by Montgomery \cite{Mo44} to characterize those
dynamical Lie algebras that lead to controllability
 as it was done by R. Brockett in the classical theory of control \cite{Br73} and more
rencently in the quantum case \cite{Sc01}, \cite{Sc02}.
 
The geometrical ideas that lie behind the techniques to study controllability in finite-dimensional systems can hardly be extended to
the infinite-dimensional case.    The knowledge we have on infinite-dimensional groups and their realizations are meager than that we have on 
finite-dimensional ones.  In particular we do not have theorems characterizing groups possessing spheres of infinite dimension as homogeneous spaces like in the finite-dimensional case.
Moreover more fundamental difficulties arise in infinite-dimensional Hilbert spaces
from the fact that in most ocassions the operators $H_0, \ldots, H_r$ are unbounded and
the construction of the dynamical Lie algebra is compromised.  As it was commented in the introduction various partial results are known that use either strong restrictions on the domains and the structure of the Lie algebras generated by the operators $H_k$ as in Clark \cite{Cl03},
 or in the structure of the Hilbert space $\mathcal{H}$ as in Block \cite{Bl06}.   In both cases criteria for the controllability of some interesting systems are obtained.

Inspired in practice a weaker notion of controllability known as approximate controllability is also used.   Given an initial state $\psi_0$,  we say that a quantum control system $H(u)$ is $\epsilon$-approximately controllable with respect to $\psi_0$, if
for any given state $\psi$, there exists a time $T$ and
a control function $u(t) \in \mathcal{U}$ such that the solution $\psi(t;u)$ satisfies $|| \psi(T;u) - \psi || < \epsilon$.    If the system is $\epsilon$-approximately controllable for all $\epsilon > 0$ and for all $\psi_0$ we say that it is approximately controllable.   The notion of approximate controllability is suitable for experimental purposes where absolute precision is impossible to achieve.    Notice that a quantum system is approximately controllable if the reachable set $\mathcal{R}$ is dense in $S$.    In this sense it is noticeable the result recently obtained by 
Chambrion {\it et al} \cite{Ch08} where it is proven that  a bilinear system:
$$ \frac{d}{dt} \psi = (A + u B) \psi ,\quad \psi \in \mathcal{H},$$
$A$, $B$ skew-adjoint operators with discrete spectrum, such that the sequence of differences of eigenvalues of $A$, $\lambda_n - \lambda_{n+1}$, is rationally independent, the elements $\langle \phi_n B, \phi_m \rangle \neq 0$ and $u$ is a piecewise-constant function, is approximately controllable.   In particular if we write the harmonic oscillator eq. (\ref{harmonic}) in the form $i \dot{\psi} = (H_0 + uH_1)\psi$ and now consider
the bilinear system $\dot{\psi} = (A + uB)\psi$ with $A = -i(H_0 + \mu H_1)$ and $B = -i(H_1-\mu H_0)$ for $\mu$ an irrational real number small enough, then the conditions of the previous theorem are fulfilled and the system is approximately controllable.  

\section{Von Neumann Controllability}\label{von Neumann}

\subsection{A simple linear superposition quantum control device based on a Mach-Zehnder-Kerr interferometer}\label{simple_interf}
\begin{figure} 
\begin{center}
\includegraphics[width=5in, height=3in]{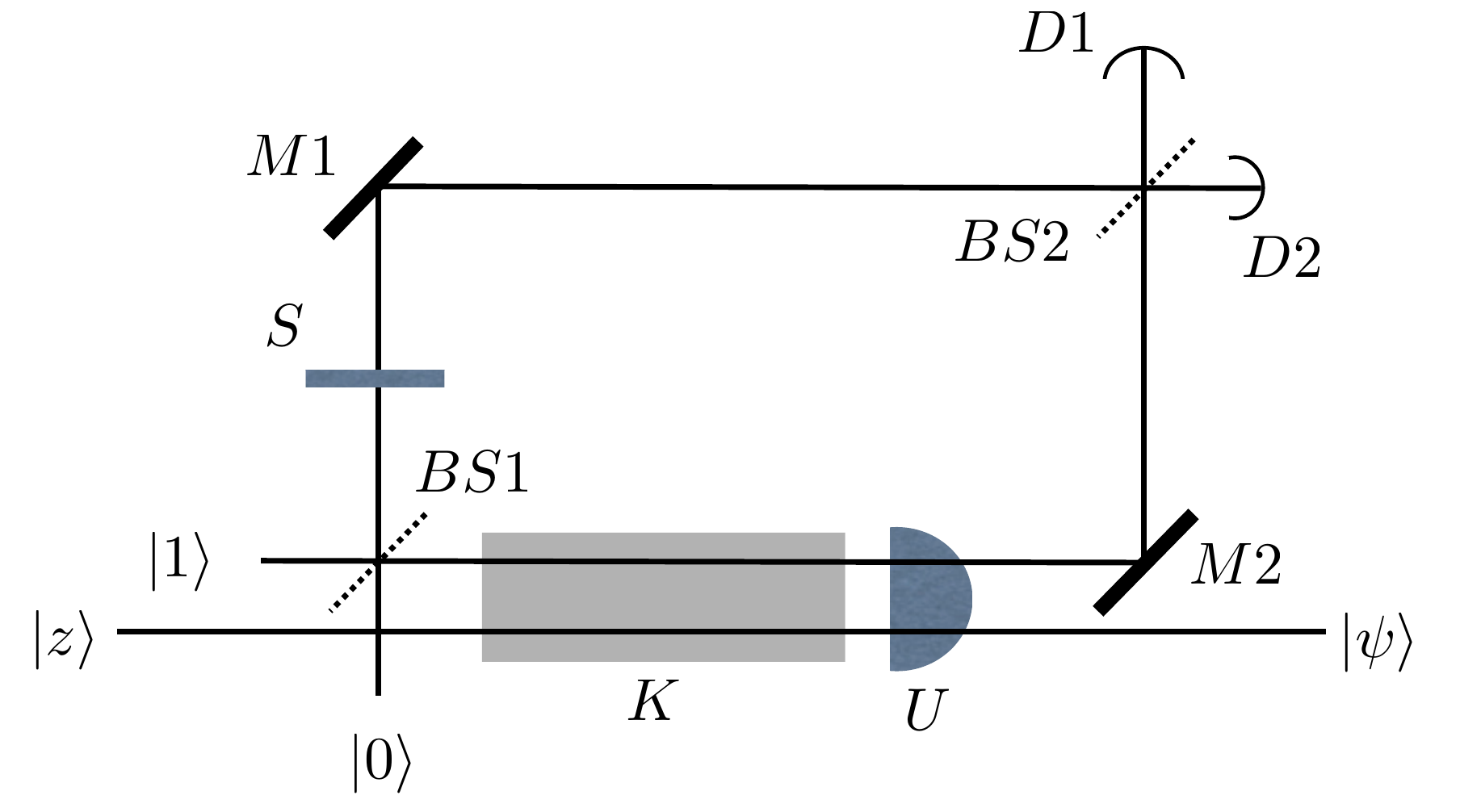} %
\caption{Scheme for the proposed method to construct controlled states by generating linearly superposed states.}
\label{kerr}
\end{center}
\end{figure}

As it was discussed in the introduction there are a number of possibilities to generate quantum superposition states, hence to try to implement a device that would allow to control states by using linear superposition.   Before embarking in the formal definition of von Neumann controllability we will discuss first one of these methods, which is based on the generation of optical macroscopic superposition states via state reduction using a Mach-Zehnder interferometer with a Kerr medium as it was discussed in \cite{Ge99}.   We will see that in addition this experiment provides effective control mechanisms for the output quantum states.   A schematic for the method is given in Fig. \ref{kerr}.      The device requires a standard Mach-Zehnder interferometer with a Kerr medium (K) in one arm of the counterclockwise path and a phase shifter (S) in the clockwise path generating a phase shift $\theta$.    

The quantum state to be controlled is a coherent state $| z \rangle$, which is a vector on the Fock Hilbert space describing the quantum states of optical laser photons.    We would like to obtain output states which are linear superpositions of the form $| \psi \rangle = (c_1 I  + c_2 U_2 + \cdots + c_N U_N ) | z \rangle$ where $c_k$ are complex numbers and $U_k$ unitary operators, $k = 1, \ldots, N$.

 Denoting by $\hat{a}, \hat{a}^\dagger$ the annihilation and creation operators on the Fock space of the laser, $|z\rangle$ is an eigenvector of $\hat{a}$ with eigenvalue $z$, $\hat{a}| z \rangle = z | z \rangle$.     To begin describing the formation of Schr\"odinger cat states we assume that single photon and vacuum states $| 1 \rangle$, $| 0 \rangle$ enter the input ports of the first beam splitter (BS1) of the interferometer, as indicated in Fig. \ref{kerr}, and that the coherent state $| z \rangle$ enters the channel that goes through the Kerr medium.
Just after the first beam splitter the state of the system is $ \frac{1}{\sqrt{2}} \left( |T\rangle + i|R \rangle  \right) | z \rangle$,
where $| R \rangle$ and $| T \rangle$ denote respectively the states in which the photon has been reflected and transmitted.    The interferometer states $|R\rangle$, $|T\rangle$ are entangled, consisting on a superposition of the states for a photon propagating along two different paths, the two arms of the interferometer.   Just before the mirrors (M1) and (M2), as a result of the Kerr interaction and the phase shifter, the state of the system is:
\begin{equation}\label{after_bs1}
\frac{1}{\sqrt{2}} \left( ie^{i\theta}|R\rangle | z \rangle +  |T \rangle  | z e^{-i\phi} \rangle \right) ,
\end{equation}
where $\phi = Kl/v$ has time units, $K$ measures the strength of the Kerr medium, $l$ is its length and $v$ is the velocity of the light in the medium.
Now, at the mirrors (M1) and (M2) both beams suffer a $\pi/2$ phase shift amounting to an overall irrelevant phase factor.  The second beam splitter, according to the superposition principle, performs the following transformations $$|R \rangle=\frac{1}{\sqrt{2}} \left(|D_2 \rangle+i|D_1 \rangle \right)\quad,\quad|T \rangle=\frac{1}{\sqrt{2}} \left(|D_1 \rangle+i|D_2 \rangle \right)$$ and the state of the system given by (\ref{after_bs1}) becomes finally:
\begin{equation} \label{after_bs2}
\frac{1}{2} \left[ |D_1\rangle \left(|z e^{-i\phi} \rangle -e^{i\theta} | z \rangle \right) +  i|D_2 \rangle \left( e^{i\theta}|z \rangle +  | z e^{-i\phi} \rangle \right)  \right] .
\end{equation}
If detector D1 and not D2 fires, then the state $|D_1\rangle $ is detected,  thus projecting the total state (\ref{after_bs2}) into the state on the Fock space of the laser given by $| \psi \rangle =  | z e^{-i\phi} \rangle-e^{i\theta} | z \rangle$.  A simple computation shows that the probabilities of obtaining this state (i.e. the probability that the state $|D_1\rangle$ is detected) is given by $P(\theta, \phi ) = \frac{1}{2}\{ 1 - \exp [- | z|^2 (1 - \cos \phi)] \cos (\theta + |z |^2 \sin \theta ) \}$.   Now, the state $|z e^{-i\phi} \rangle $ is obtained as a unitary evolution of state $|z \rangle$ with respect to the harmonic oscillator hamiltonian $H_0 = \hat{a}^\dagger \hat{a} + 1/2$ as $e^{i\phi/2}e^{-i\phi H_0} |z \rangle = |z e^{-i\phi} \rangle$.   Denoting the unitary operator $e^{i\phi/2}e^{-i\phi H_0}$ by $U_\phi$ we get that the output state could be written as $|\psi \rangle = (e^{i\theta} I + U_\phi)|z\rangle$ which has the form that we were looking for.

Moreover, if we insert a usual quantum gate $U$ between the Kerr medium and the mirror (M2), see Fig. \ref{kerr}, coupling the interferometer and the laser channel,  of the form: 

$$U (|R\rangle | z \rangle) = \cos \alpha |R \rangle U_1 |z \rangle + \sin \alpha | T\rangle U_2 | z \rangle ,$$ 
$$U (|T\rangle | z \rangle) = -\sin \alpha |R \rangle U_1 |z \rangle + \cos \alpha | T\rangle U_2 | z \rangle ,$$ 
where $U_k$, $k = 1,2$ are unitary operators on Fock space,  we get that the final output state $| \psi \rangle$ above becomes:

\begin{equation}\label{gen_cat}  
| \psi \rangle = (e^{i\theta} I + i \cos \alpha U_1 U_\phi - \sin \alpha U_2 U_\phi )|z\rangle .
\end{equation}
Now the parameter $\phi$, which is proportional to the length of the Kerr medium, and the pase shift $\theta$ act as control parameters, that together with the quantum gates $U_k$ allow to generate a superposition of quantum states each one evolved by a different unitary operator.

\subsection{On the general notion of linear superposition quantum controllability}

We assume now that we have an affine-linear quantum control system $H(u) = H_0 + \sum_{k = 1}^r u_k H_k$ such that the
family of self-adjoint operators $H_0, \ldots, H_r$ generate a finite-dimensional Lie algebra $\mathfrak{g}$ (i.e. the skew symmetric operators $iH_k$ are generators of the Lie algebra $\mathfrak{g}$).   In such a case the dynamical Lie group $G$ corresponding
to the dynamical Lie algebra $\mathfrak{g}$ is represented unitarily on the Hilbert space $\mathcal{H}$ of the system.    If we denote by $A_l$ a basis of the Lie algebra $\mathfrak{g}$, then any element $g$ on the group $G$ can be written as: 
\begin{equation}\label{group_elem}
g = \prod_{r<\infty}\exp{\tau_{l_r}A_{l_r}},
\end{equation}
for some family of real numbers $\tau_{l_r}$. The solutions of Schr\"odinger's equation (\ref{schro}) for piecewise constant control functions $u_k(t)$ consist of piecewise differential curves $\psi(t;u)$ of the form:
$$ \psi(t;u) =   U(t,t_s)U(t_s,t_{s-1}) \cdots U(t_2,t_1)U(t_1,t_0) \psi_0,$$
where $u(t)$ is defined on the interval $[t_0,T]$ with discontinuity points  $t_1 < t_2 < \cdots < t_f$, $t_s \leq t < t_{s+1}$, and $U(t_j,t_{j-1})$  are unitary operators of the form:
\begin{equation} \label{unit_group}
U(t',t) =  \prod_{j<\infty}e^{\tau_{l_j}\hat{A}_{l_j}} ,
\end{equation}
with $\hat{A}_{l_j}$ the skew-hermitian operators realizing the basis elements $A_{l_j}$ (notice that the operators $\hat{A}_l$ realizing the basis $A_l$  are linear combinations of commutators of various orders of the elements $iH_k$).   Mimicking equation (\ref{gen_cat}) for families of unitary operators of the form (\ref{unit_group}), we propose the following natural notion of controllability:

 \begin{definition}  An affine quantum control system with a finite-dimensional dynamical Lie algebra $\mathfrak{g}$, will be said to be von Neumann controllable with respect to the state $\psi_0$  if for any state $\psi_1$ and any $\epsilon > 0$, there exists a family of coefficients $c_k$, $k=1, \ldots, N$ ($N$ depending on $\epsilon$) and for each $k$ a family of times $t_{k_l}$, $l = 1,\ldots, M_k$, such that:
 $$ || \psi_1 - \sum_{k=1}^N c_k \prod_{l=1}^{M_k} e^{t_{k_l} \hat{A}_{k_l}} \psi_0 || < \epsilon ,$$
 where $A_j$ is a basis of the dynamical Lie algebra of the system.  Moreover we will say that the system is von Neumann controllable if it is von Neumann controllable with respect to any state $\psi_0$.
 \end{definition}
  
The notion of von Neumann controllability is a
notion of approximate controllability.  It is clear from the definitions and the discussion above that an approximately controllable quantum system is von Neumann controllable.  However the contrary is not necessarily true as it will be discussed later on.

Let us insist, that in sharp contrast with the notion of pure state controllability, von Neumann controllability allows for the use of linear superposition of states to reach the target states.    The algebra of operators on a Hilbert space generated by operators of the form (\ref{unit_group}) is called a von Neumann algebra and this is the reason of the name chosen for the above notion of controllability.

\section{Von Neumann controllability and representation theory}\label{representation}
Its now time to discuss the relation of von Neumann controllability with the unitary representations of the dynamical group of the system.

We should notice that for each element $g$ of the dynamical group of the form (\ref{group_elem}), we have a unitary operator associated 
\begin{equation}\label{unitary}
U(g) = \prod_{s<\infty}e^{t_{l_s}\hat{A}_{l_s}}.
\end{equation}
Thus the map $U(g)$ provides a unitary representation of the group $G$.    In the particular case that the operators $iH_k$ do define a basis for a Lie algebra, i.e. they are independent and satisfy commutation relations of the form 
$$ [H_k,H_j ] = iC_{kj}^l H_l , \quad k.j,l = 0, \ldots, r ,$$
the group $G$ is represented by the unitary operators $U_k (t) = e^{itH_k}$.   It is clear then that the unitary representation $U\colon G \to U(\mathcal{H})$ is strongly continuous because for any vector $\psi\in \mathcal{H}$ the map $G \to \mathcal{H}$ given by $g\mapsto U(g)\psi$ is continuous.

Thus in the study of controllability of affine quantum systems we are led naturally to consider its relation with the theory
of unitary representations of groups.   In fact a first simple observation in this sense is the following:

\begin{theorem}\label{control_irr}  Let us consider an affine-linear quantum control system $H(u) = H_0 + \sum_{k=1}^r u_k H_k$ such that its dynamical group $G$ is a finite-dimensional Lie group.   Then a necessary condition for the pure state controllability
of the system is that the unitary representation $U$ of the dynamical group is irreducible.
\end{theorem}

{\noindent {\em Proof:}}   Denoting by $U(g)$ the unitary representation of the dynamical Lie group $G$ on $\mathcal{H}$, it is clear that if $U$ is not irreducible, then there exists a proper invariant closed subspace $W\subset \mathcal{H}$.    Notice that the unitary operators $U(g)$ leave invariant the pure states of the system $U(g) S \subset S$, for all $g\in G$.  Therefor it is clear that $W \cap S$ will be an invariant subset of $S$ strictly contained on it.  Hence, there will exist vectors $\psi \in S$ such that $\psi \notin W\cap S$.  Such vectors will not be reachable from any vector on $W$. \hfill $\Box$

\bigskip

Given a unitary representation $U$ of a Lie group $G$ on a Hilbert space $\mathcal{H}$ it is natural to
consider for any vector $\psi\in \mathcal{H}$ its ``linear orbit
'', this is the linear closure of all vectors of the form $U(g) \psi$, $g \in G$, or in other words the linear closure of the actual orbit of the vector $\psi$ under the action of the group $G$.   Such linear subspace will be denoted by $\mathcal{H}_\psi^U$, i.e.
$$\mathcal{H}_\psi^U = \overline{\mbox{span} \{ U(g)\psi \mid g \in G \}} .$$

 A vector $\psi$ is called cyclic if $\mathcal{H}_\psi^U = \mathcal{H}$ and the corresponding representation is said to be cyclic.  
 Hence it is clear that the quantum system of Thm.  \ref{control_irr} will be von Neumann controllable with respect to the state $\psi_0$ if $\psi_0$ is a cyclic vector for the unitary representation $U$.
 
  It is also clear that  any non-zero vector on the vector space of an irreducible representation $U$ is cyclic.   Conversely if all vectors of an unitary representation are cyclic the representation is irreducible.  Taking advantage of this we can state
the following theorem that provide us with a sharp criterium to determine when an affine quantum system is going to be von Neumann controllable.

\begin{theorem}  An
affine-linear quantum system  $H(u) = H_0 + \sum_{k=1}^r u_kH_k$ with finite-dimensional dynamical Lie algebra $\mathfrak{g}$ is von Neumann controllable if and only if the unitary representation $U$ of its dynamical Lie group $G$  is irreducible.

\end{theorem}

{\noindent {\em Proof:}}   If the unitary representation $U(g)$ is irreducible, every nonzero vector $\psi_0$ is cyclic.  Then because  $\mathcal{H} = \overline{\mbox{span} \{ U(g)\psi \mid g \in G \}}$, the set of finite linear combinations of vectors $U(g)\psi$ is dense on $\mathcal{H}$.  Hence for any vector $\psi_1$ and for any $\epsilon > 0$ there exists a family of elements $g_i$ and constants $c_i$, $i=1, \ldots, m$ such that 
$$ || \psi_1 - \sum_{i=1}^m c_i U(g_i) \psi_0 || < \epsilon .$$
Moreover, any unitary operator of the representation $U(g)$ can be written as in eq. (\ref{unitary}).  Substituting back in the previous formula we find the expression defining von Neumann controllability.

Conversely, if the quantum system is von Neumann controllable, then for any nonzero vector $\psi$ we have that $\mathcal{H}= \overline{\mbox{span} \{ U(g)\psi \mid g \in G \}}$, thus any nonzero vector is a cyclic vector for the unitary representation $U(g)$.  Hence the representation is irreducible. \hfill $\Box$

\bigskip

However we should stress that reducible representations of groups can contain cyclic vectors.  This opens the possibility of von Neumann control with respect to a given vector even if the unitary representation of the group is reducible, obtaining in this way controllability with respect to a given vector by using just a few controls and providing a way of improving the efficiency of the control methods.   For instance, if we consider a spin $1/2$ system, it trivially supports an irreducible unitary representation of the group $SU(2)$, however such representation is reducible for any $U(1)$ subgroup of $SU(2)$.    Now if we select a $U(1)$ subgroup of $SU(2)$ and consider the North-South axis defined by it on $S^2$, then any non-equatorial vector is a cyclic vector with respect to the (reducible) unitary representation defined by the given subgroup $U(1)$ on $\mathbb{C}^2$. This system is then von Neumann controllable for a wide subspace of its Hilbert space but the control group is much smaller than the  one  needed to achieve pure state controllability, i.e., $SU(2)$. Of course the question of whether a given initial state is cyclic or not in a reducible representation is still open and we leave this issue for future work. \\

\section{Some examples and applications}\label{examples}

\subsection{Von Neumann control of coherent states on a Mach-Zehnder-Kerr interferometer}

We are ready now to finish the discussion of the controllability of coherent states on a Mach-Zehnder-Kerr interferometer started in section \ref{von Neumann}.
In order to do that let us consider again the example of the controlled quantum harmonic oscillator (\ref{harmonic}).  The controlled harmonic oscillator has a dynamical algebra generated by the self-adjoint operators 
$$H_0 = -\frac{1}{2}\frac{\partial^2}{\partial q^2} + \frac{1}{2}q^2, \quad H_1 = -q ,$$
on the Hilbert space $L^2(\mathbb{R})$.  Because of the commutation relations:
$$[H_0,H_1] = iH_2, \quad [H_0,H_2] = -iH_1, \quad [H_1,H_2] = -i I,$$
with $H_2 = i\frac{\partial}{\partial q}$, the dynamical Lie algebra of this system is a 4-dimensional Lie algebra called the oscillator algebra and the Lie group defined by it
the oscillator group.   Such group is a 4-dimensional solvable Lie group whose unitary representations
where thoroughly
studied for the first time in a remarkable paper by Streater where a detailed comparative
analysis of the construction of its unitary irreducible representations using both Mackey's theory and 
Kirillov's coadjoint orbit quantization construction was done \cite{St67}.    Hence according to Streater's classification
the previous representation of the oscillator group is irreducible (in fact it is essentially the only one that has physical sense)
and the system is von Neumann controllable.   However we already know that it is not pure state controllable.
This shows that the notion of von Neumann controllability
is different and strictly weaker than the notion of pure state controllability.
We also know that the controlled quantum harmonic oscillator
is approximately controllable.  Thus, in this case, von Neumann controllability and approximate pure state controllability are
equivalent.  

We can apply this result to the example of a Mach-Zehnder-Kerr interferometer if in addition to the harmonic oscillator hamiltonian $H_0= \hat{a}^\dagger \hat{a} + 1/2$ that describes the interaction of the Kerr medium with the coherent state, we introduce another medium whose effective hamiltonian has a term $H_1=-q=-(\hat{a}^\dagger + \hat{a})/\sqrt2$.   Such term will replace the
quantum gate given by the unitary operator $U$ and therefore the evolution of the system is governed now by the oscillator algebra above and the
system is controllable in the von Neumann sense for all initial states.   The Jaynes-Cummings models provide many different realizations of such interaction term \cite{Sh93}, \cite{Jo92}, as well as other possibilities for interaction terms that could be easily implemented in a concrete experimental setting.    

\subsection{Compact Lie groups}  

Once the connection between von Neumann controllability and the theory of unitary representations has been established, we can start a systematic discussion of the controllability of quantum systems by discussing the irreducible unitary representations of their dynamical Lie groups.   Unfortunately this programme cannot be fully carried on because there is not a complete theory of irreducible representations for all Lie groups. 
However we have a fairly well developed theory for various families of Lie groups.   We will not pretend here to cover such broad scenario and concentrate on two opposite cases which are both
extremely useful, on physical and mathematical grounds, compact and nilpotent Lie groups.  For both families of Lie groups their re\-presentations are very well-known.  The theory of representations of compact Lie groups was developed at the begining of the XX century by Weyl together with the creation of the mathematical foundations of quantum mechanics.  The theory of unitary representations of nilpotent Lie groups took some more time to mature and Kirillov in 1960 prove the celebrated theorem stablishing a one-to-one correspondence between unitary irreducible representations of nilpotent Lie groups and coadjoint orbits in the dual of the Lie algebra of the group, marking the beginning of the ``method of orbits'' leading to the modern paradigm of geometric quantization.

The Peter-Weyl theorem establishes
that the set $\hat{G}$ of equivalence classes of irreducible unitary representations of a compact Lie group $G$  is a countable discrete set. Morover that the regular representation of the group, i.e. the Hilbert space $L^2(G, \mu_G)$ where $\mu_G$ is the Haar measure of the group, decomposes as:
$$ L^2(G, \mu_G) = \bigoplus_{\alpha \in \hat{G}} \mathcal{H}_\alpha\otimes \mathcal{H}_\alpha' ,$$
where the finite-dimensional $\mathcal{H}_\alpha$ is the support  space of the irreducible unitary representation labelled by $\alpha$ and $\mathcal{H}_\alpha'$ its dual space.  If we denote by $d_\alpha$ the dimension of $\mathcal{H}_\alpha$, the previous formula also tells that the multplicity of the irreducible representation supported at $\mathcal{H}_\alpha$ on the regular representation is $d_\alpha$.   Hence, if we denote by $R$ the regular representation, this is $(R(g)\psi)(g') = \psi(gg')$, $g,g' \in G$, $\psi\in L^2(G,\mu_G)$, we have:
$$ R = \bigoplus_{\alpha \in \hat{G}} d_\alpha U_\alpha ,$$
where $U_\alpha \colon G \to U(\mathcal{H}_\alpha)$ denotes the corresponding irreducible unitary representation of $G$.
Hence for any affine-linear quantum system defining a unitary representation of a compact Lie group $G$, the system will be von Neumann controllable iff $\mathcal{H} = \mathcal{H}_\alpha$ for some $\alpha \in \hat{G}$.    Moreover the system in general will not be controllable unless $\dim G \geq d_\alpha^2$.  In fact if the system is von Neumann controllable and hence it supports the irreducible representation $\mathcal{H}_\alpha$ then the map $U\colon G \to U(\mathcal{H}_\alpha)$ maps $G$ as a subgroup of $U(\mathcal{H}_\alpha ) \cong U(d_\alpha)$.  Thus the system will be controllable if $G$ is mapped surjectively on $U(d_\alpha )$ and this will only occur if $\dim G \geq d_\alpha^2$.

\subsection{Nilpotent Lie groups}  According to Kirillov's theorem any unitary irreducible representation of a nilpotent group $G$ can be constructed by induction of one-dimensional representations of suitable chosen subgroups $H$ of $G$.    Consider the dual $\mathfrak{g}^*$ of the Lie algebra $\mathfrak{g}$ of the Lie group $G$.   Let $\mu  \in\mathfrak{g}^*$ and $\mathcal{O}_\mu$ the coadjoint orbit through it, this is, $\mathcal{O}_\mu = \{ \mbox{Ad}_g^* \mu \mid g \in G\}$, where $\mbox{Ad}_g^*$ denotes the adjoint of the adjoint action $\mbox{Ad}_g$ of $G$ on $\mathfrak{g}$.
Let $H = G_\mu$ be the isotropy group of $\mu$, i.e. $G_\mu = \{ g \in G \mid \mbox{Ad}_g^* \mu = \mu \}$ and $\mathfrak{h}$ its Lie algebra.  Let us consider the character of $H$ defined by $\mu$, this is, $\chi_\mu (\exp{\zeta}) = e^{i\langle \mu, \zeta \rangle }$.   The character $\chi_\mu$ defines a one-dimensional unitary representation of the subgroup $H$ of $G$.   Then the unitary representation of $G$ obtained by induction from $\chi_\mu$, $U^{\chi_\mu}$ is irreducible.  Recall that $U^{\chi_\mu}$ is an infinite-dimensional representation whose support space is the Hilbert space of square integrable functions $\psi \colon G \to \mathbb{C}$ such that $\psi (hg) = \chi_\mu(h)^* \psi (g) $, $g \in G$, $h \in H$;  moreover, $U^{\chi_\mu}(g) \psi(g') = \psi (gg')$.    According to this construction if an affine-linear quantum control system defining a nilpotent dynamical Lie group is von Neumann controllable, i.e. it defines an irreducible representation of $G$, it is infinite-dimensional,
 hence it will not be pure state controllable.  Notice that the only states reachable from a given one $\psi_0$ will be those translated by elements of the group, thus conforming a finite-dimensional orbit on state space.     It is not obvious however whether the given system is going to be approximately controllable.  Let us consider the simple example of a controlled free particle on $\mathbb{R}$:
\begin{equation}\label{free}
i \frac{\partial \psi}{\partial t} = - \frac{1}{2} \frac{\partial^2 \psi}{\partial q^2}  -u(t) q\psi = (H_0 + uH_1) \psi ,
\end{equation}
Then the dynamical Lie algebra
generated by $H_0$ and $H_1$ is a 4-dimensional nilpotent Lie algebra that contains the Heisenberg algebra.  The Hilbert space $L^2(\mathbb{R})$ supports a (essentially unique) unitary representation of the Heisenberg group according with von Neumann's theorem, thus it supports an irreducible unitary representation of the dynamical Lie group, hence the system will be von Neumann controllable. We cannot apply Chambrion's theorem to it because the spectrum of the operator $H_0$ is not discrete. However it is easy to check that the controlled free particle above is not approximately controllable.

\section{Conclusions and outlook}

We have presented a new notion of control, von Neumann controllability, for quantum states that extends the usual notion of state controllability using the linear superposition principle of Quantum Mechanics.   The output states are a linear superposition of states each obtained by evolving the input state using the control hamiltonian with various control functions.    A quantum system will be von Neumann controllable with respect to a given initial state if such state is a cyclic vector for the unitary representation of the dynamical group generated by the control hamiltonian, and the system will be said von Neumann controllable if it is controllable in this sense for all initial states.    The notion of von Neumann controllability is equivalent to the irreducibility of such unitary representation if and only if the system is von Neumann controllable with respect to all unitary vectors.  
The characterization of states such that the system is von Neumann controllable with respect to them for a given reducible representation of the dynamical group is an open question that will be addressed in forthcoming works.    

We have also shown that the notion of von Neumann controllability is weaker than the notion of approximately pure state controllability because there are von Neumann controllable systems (those supporting irreducible representations of the dynamical group) that cannot be approximately controllable because the dimension of the representation space is much larger than the dimension of the group.    On the other hand von Neumann controllability can be applied without further difficulties both to finite and infinite dimensional systems.     We have started the analysis of some simple infinite dimensional systems which are von Neumann controllable by using some rudiments of the theory of irreducible unitary representations of nilpotent Lie groups.  The problem becomes much  more interesting when considering solvable groups and we leave it for further analysis.

\ack This work was partially supported by MEC grant MTM2007-62478 and SIMUMAT project.  We would also thank the comments and questions raised by the referees that have helped to shape the final form of this paper.

\end{document}